\documentclass[10pt]{article}

 \usepackage{amsmath,amsthm,amssymb}
 \usepackage{makeidx,epsfig,lscape}
 \usepackage{color,colortbl}
 \usepackage{fancyhdr}
 \usepackage{xcolor,pict2e}

 \renewcommand{\headrulewidth}{0pt}
 \renewcommand{\footrulewidth}{0.5pt}

 \definecolor{myaqua}{rgb}{0.0,0.5,0.55}
 \definecolor{lightaqua}{rgb}{0.75,0.95,0.95}

 \usepackage[colorlinks = true,
            linkcolor = myaqua,
            urlcolor  = blue,
            citecolor = myaqua]{hyperref}

\usepackage{caption}
\usepackage{floatrow}
\def\lin#1#2{\textcolor[rgb]{0.6,0.6,0.6}{\vspace*{#1mm} \hrule
   height 3 pt \vspace*{#2mm}}}

\def\bt{\begin{tabular}}
\def\et{\end{tabular}}
\def\and{\mbox{ and }}

\def\1{{\bf 1}}

 \def\sectionn#1{\refstepcounter{section}{\color{myaqua}

 \vskip 6mm

 \noindent\Large\bf\thesection. #1}

 \vskip 3mm}
 \def\subsectionn#1{\refstepcounter{subsection}{\color{myaqua}

 \vskip 5mm

 \noindent\large\bf\thesubsection. #1}

 \vskip 2mm}

 \def\boxx#1#2#3#4#5{
 {\linethickness{#4pt}\put(#1,#5){\color{myaqua}{\line(1,0){#3}}}}
 \multiput(#1,#2)(0,#4){2}{\line(1,0){#3}}
 \multiput(#1,#2)(#3,0){2}{\line(0,1){#4}}
  }

\begin{document}

 \fancyhead[L]{\hspace*{-13mm}
 \bt{l}{\bf Journal of  Modern Physics, 2014, 5, 1945-1957}\\
 Published Online November 2014 in SciRes.
 \href{http://www.scirp.org/journal/jmp}{\color{blue}{\underline{\smash{http://www.scirp.org/journal/jmp}}}} \\
 \href{http://dx.doi.org/10.4236/jmp.2014.517189}{\color{blue}{\underline{\smash{http://dx.doi.org/10.4236/jmp.2014.517189}}}} \\
 \et}
 \fancyhead[R]{\includegraphics{pic1.ps}}

 $\mbox{ }$

 \vskip 12mm

{ 

{\noindent{\huge\bf\color{myaqua}
  On the Dynamics of the Ensemble of Particles \\[2mm] in the Thermodynamic Model of Gravity}}
%
\\[6mm]
{\large\bf Merab Gogberashvili$^{1,2}$}}
\\[2mm]
{ 
 $^1$ Andronikashvili Institute of Physics, Tbilisi, Georgia\\
 $^2$ Javakhishvili State University, Tbilisi, Georgia\\
Email: \href{mailto:gogber@gmail.com}{\color{blue}{\underline{\smash{gogber@gmail.com}}}}}
 \\[4mm]
Received 20 September 2014; revised 18 October 2014; accepted 14 November 2014
 \\[4mm]
Copyright \copyright \ 2014 by author(s) and Scientific Research Publishing Inc. \\
This work is licensed under the Creative Commons Attribution International License (CC BY). \\
\href{http://creativecommons.org/licenses/by/4.0/}{\color{blue}{\underline{\smash{http://creativecommons.org/licenses/by/4.0/}}}}\\
 \includegraphics{pic2.ps}

\lin{5}{7}

 { 
 {\noindent{\large\bf\color{myaqua} Abstract}{\bf \\[3mm]
 \textup{Within the thermodynamic model of gravity the dark energy is identified with the energy of collective gravitational interactions of all particles in the universe, which is missing in the standard treatments. For the model-universe we estimate the radiation, baryon and dark energy densities and obtain the values which are close to the current observations. It is shown that total gravitational potential of a particle from the world ensemble is a scale dependent quantity and its value is twice of Newtonian potential. The Einstein-Infeld-Hoffmann approximation to general relativity was used to show that the acceleration of a particle from the world ensemble can be considered as a relative quantity when the universe is described by the flat cosmological model.
 }}}
 \\[4mm]
 {\noindent{\large\bf\color{myaqua} Keywords}{\bf \\[3mm]
 Thermodynamic gravity; Cosmological parameters; Einstein-Infeld-Hoffmann formalism
}

 \fancyfoot[L]{{\noindent{\color{myaqua}{\bf How to cite this paper:}} M. Gogberashvili, {\it On the Dynamics of the Ensemble of Particles in the Thermo-
dynamic Model of Gravity}, J. of Mod. Phys. {\bf 5} (2014) 1945-1957. {\underline{\smash{http://dx.doi.org/10.4236/jmp.2014.517189}}}}}

\lin{3}{1}


\sectionn{Introduction}

{ \fontfamily{times}\selectfont
 \noindent 
 To construct a physical theory it is convenient to use inertial reference frames. The crucial question is: What are the inertial frames? How are they found? The precise acceleration of the Earth relative to the universe as a whole is quite difficult to measure. However, now it is possible to find quite accurately the absolute velocity of the Earth with respect to the distant stars or cosmic microwave background. Such measurements reveal that 'universal' reference frames, measured by these two different methods, coincide and constant velocities with respect to the universe seems to correspond to inertial frames \cite{Bar}. The known observational evidence that our universe plays a crucial role in the definition of the inertial frames is the fact that the universe as a whole does not rotate \cite{rindler}.

The possibility of identifying the absolute reference frame of the universe can be understand as an observational verification of Mach's principle \cite{Cheng}. Machian approach, which assume profound relations between the local and the global physics, have been subject to much discussion and speculation (for some resent studies, see \cite{Bar-Fis}). Several aspects of Mach's principle are discussed in the literature \cite{Mach}. One such assumption that the inertial frames are determined by the distant masses has been successfully incorporated in Einstein's theory of gravity \cite{Wei}. Recent results on the Gravity probe B experiment \cite{ProbeB} verified for the first time the Machian frame dragging prediction of General Relativity \cite{FrameDrag}.

Machian models usually assume that inertia of a particle with the mass $m$ is determined by the gravitational field of the whole universe and the particles total energy (inertial + gravitational) at rest with respect to the universe is zero \cite{Mach},
\begin{equation} \label{balance}
mc^2 + m\Phi =0~,
\end{equation}
where $\Phi$ denotes the gravitational potential of the whole universe acting on the particle. Energy balance conditions like (\ref{balance}) mean that the total energy of any object in the universe vanishes, with the gravitational energy of the interaction with the universe assumed to be negative and all other forms of energy be positive \cite{FMW}. Hence, the total energy of the whole universe vanishes and it can emerge without violation of the energy conservation. This is the point of view which appears to be preferable in cosmology \cite{FMW,Haw}.

\renewcommand{\headrulewidth}{0.5pt}
\renewcommand{\footrulewidth}{0pt}

 \pagestyle{fancy}
 \fancyfoot{}
 \fancyhead{} 
 \fancyhf{}
 \fancyhead[RO]{\leavevmode \put(-90,0){\color{myaqua}Merab Gogberashvili} \boxx{15}{-10}{10}{50}{15} }
 \fancyhead[LE]{\leavevmode \put(0,0){\color{myaqua}Merab Gogberashvili}  \boxx{-45}{-10}{10}{50}{15} }
 \fancyfoot[C]{\leavevmode
 \put(0,0){\color{lightaqua}\circle*{34}}
 \put(0,0){\color{myaqua}\circle{34}}
 \put(-2.5,-3){\color{myaqua}\thepage}}

 \renewcommand{\headrule}{\hbox to\headwidth{\color{myaqua}\leaders\hrule height \headrulewidth\hfill}}

It is known that the usual interpretation of Mach's principle that inertia is a relative quantity and is not attributed to an object, leads to the anisotropy of the rest mass of a particles due to the influence of nearby massive objects (like the Galaxy) \cite{Bar-Fis}. This possibility has been ruled out by experiments \cite{exp}: in agreement with Einstein's equivalence principle inertia is observed to be highly isotropic \cite{Wei}.

As an attempt to address these problems a thermodynamic model of gravity, where the universe is considered as the statistical ensemble of all gravitationally interacting particles inside the horizon, was proposed in \cite{Gog}. The naive Machian conjecture that the mass parameter can be altered by "distant stars" is based on the assumption that kinematics, or space-time, exists separately from the dynamics of masses and is independent of the surrounding universe. The lesson of General Relativity has been that the description of space-time geometry, and hence the kinematics itself, depend on the distribution of matter. Therefore, Mach's principle should be considered at the level of elementary particles and in terms of more fundamental quantities, such as action or information transfer, than the mass parameter. Quantum mechanics says that physical systems isolated from the world ensemble of elementary particles do not exist. Since the number of particles in the universe is huge, the influence of world ensemble of particles on local physics, or Mach's principle in its utmost generality, does not lead to the observable anisotropy of the masses in thermodynamic approach.

According to the thermodynamic model \cite{Gog}, a more appropriate way to describe gravity is in terms of changes of the thermodynamic properties in the world ensemble, such as temperature or entropy, rather than in terms of space-time geometry, which can be derived as an emerging effective description (see \cite{Thermo,Pad,Masud} and citations therein). It was demonstrated that the model is compatible with the existing field-theoretical descriptions, as the relativistic and quantum properties are emerging from the properties of the world ensemble.

Within the model \cite{Gog} fundamental physical constants represent collective characteristics of the world ensemble and thus are related to the cosmological parameters, in the spirit of \cite{large}. For instance, the Planck's action quantum is identified with the amount of action of an average member of the world ensemble:
\begin{equation} \label{A}
A = - m c \lambda =: - 2\pi\hbar ~,
\end{equation}
were $\lambda$ is a characteristic length of a particle when it can be considered as an oscillator, i.e. its Compton wavelength. Also the energy balance condition (\ref{Phi}) written as the Schwarzschild-like relation,
\begin{equation} \label{Phi}
c^2 = -\Phi = \frac{2GM}{R}~,
\end{equation}
where $M$ and $R$ correspond to the total mass and the radius of the universe, can be interpreted as the definition of the universal constant of the speed of light. The collective gravitational potential of all particles in the universe, $\Phi$, acting on each member of the ensemble, and thus $c$, can be regarded as constants according to the cosmological principle (the universe is isotropic and homogeneous at the scale $R$). Possible relations of cosmological parameters with the properties of the world ensemble of particles are considered in Section $2$.

If we interpret $c$ as a cosmological parameter, then $M$ in (\ref{Phi}) cannot be interpreted as ordinary gravitational mass of a classical object. It will be clear if we replace $R$ in (\ref{Phi}) by the time depended Hubble parameter:
\begin{equation} \label{Hubble}
H = \frac cR ~.
\end{equation}
To avoid variations of fundamental physical constants, $c$ or $G$, which is ruled out by experiments \cite{Variation}, we need to introduce the unrealistic condition $M/R = const$. The relation (\ref{Hubble}) follows from another "cosmological definition" of the speed of light:
\begin{equation}
c^2 = H^2R^2 \Omega~,
\end{equation}
where we introduced the total energy density of the flat cosmological model:
\begin{equation} \label{Omega}
\Omega = \Omega_M + \Omega_\Lambda = 1~,
\end{equation}
which is the only cosmological model for which $\Omega$ (the sum of the matter, $\Omega_M$, and the dark energy, $\Omega_\Lambda$, densities) is an unvaried quantity. Note also the presence of the factor $2$ in (\ref{Phi}), which distinguishes $\Phi$ from the standard Newtonian gravitational potential considered in \cite{Mach}. The reason is that the parameter,
\begin{equation} \label{M}
M = \frac{c^3}{2 GH} ~,
\end{equation}
should take into account the total matter content of the universe and not only the ordinary matter for which the classical Newton's law is written. This point will be discussed in details in Section $3$.

In Section $4$ we demonstrate that the inertia of a particle can be related to the total energy content of the universe. In standard physics acceleration is absolute in origin and all forces arise from close sources. We want to show that the acceleration can be considered as a relative quantity and Newton's second law can be written in the form:
\begin{equation}\label{F}
m \vec{a} - m \vec{a}_u = \vec{F}~ .
\end{equation}
In this formula the reactive acceleration:
\begin{equation}
\vec{a}_u=\frac{d\vec{u}}{dt}~,
\end{equation}
appears due to the non-local forces of the surrounding universe and $\vec{u}$ is an overall velocity of the universe relative to the origin of the coordinate system. So only the acceleration relative to the universe as a whole, $(\vec{a} - \vec{a}_u)$, is what matters in the equation of motion (\ref{F}). To demonstrate the validity of (\ref{F}) we shall use Einstein-Infeld-Hoffmann approximation to general relativity written for the world ensemble for the flat cosmological model.}

\renewcommand{\headrulewidth}{0.5pt}
\renewcommand{\footrulewidth}{0pt}

 \pagestyle{fancy}
 \fancyfoot{}
 \fancyhead{} 
 \fancyhf{}
 \fancyhead[RO]{\leavevmode \put(-90,0){\color{myaqua}Merab Gogberashvili} \boxx{15}{-10}{10}{50}{15} }
 \fancyhead[LE]{\leavevmode \put(0,0){\color{myaqua}Merab Gogberashvili}  \boxx{-45}{-10}{10}{50}{15} }
 \fancyfoot[C]{\leavevmode
 \put(0,0){\color{lightaqua}\circle*{34}}
 \put(0,0){\color{myaqua}\circle{34}}
 \put(-2.5,-3){\color{myaqua}\thepage}}

 \renewcommand{\headrule}{\hbox to\headwidth{\color{myaqua}\leaders\hrule height \headrulewidth\hfill}}

\sectionn{Estimation of cosmological parameters}

{ \fontfamily{times}\selectfont
 \noindent
At first let us show that realistic values of cosmological parameters can be obtained for the simplest model-universe of radius $R$ which contains the ensemble of $N$ identical particles of the mass $m$. The characteristic length of the world ensemble, or the size of the model-universe, can be estimated as:
\begin{equation}\label{lambda}
2R = \lambda N ~,
\end{equation}
were $\lambda$ is a characteristic length (the Compton wavelength) of a particle. Since each particle interacts with all other $(N-1)$ particles, and the mean separation of the interacting pairs is $R/2$, the total Newtonian energy of the ensemble consists of $N(N-1)/2$ terms of magnitude $\approx G m^2/R$. Then the energy of single particle which interacts with the total gravitational potential of the universe $\Phi$ is given by:
\begin{equation} \label{E}
E \approx \frac{N^2}{2} \frac {Gm^2}{R} ~.
\end{equation}
Correspondingly, according to the relation like (\ref{Phi}), the gravitational mass of the world ensemble is a quadratic function of the number of particles:
\begin{equation} \label{M=N2m}
M_G \approx \frac 14 N^2 m ~.
\end{equation}

In the model \cite{Gog} the gravitational energy of collective interactions of all particles is identified with the dark energy of the universe. Then the dark energy density can be expressed as the ratio of the total gravitational to the total mass of the universe:
\begin{equation} \label{Edark}
\Omega_\Lambda := \frac{M_G}{M} \approx \frac{N^2m}{4M} ~.
\end{equation}
Indeed, under the above identification the energy balance condition (\ref{balance}), written for all $N$ particles in the universe, is equivalent to the equation of state of the dark energy:
\begin{equation}
  \frac p\rho = -1~.
\end{equation}
In our model the appearance of the "exotic negative pressure $p$" has a natural explanation as the consequence of the negative collective gravitational potential $\Phi$ of the whole universe. Besides, the assumption (\ref{Edark}) also naturally answers the question as to why the density of dark energy is so close to the critical density. The standard cosmology, where $\Omega_\Lambda$ is related to the cosmological constant $\Lambda$, offers no reasonable explanation of this observational fact and the attempt to relate $\Lambda$ to the quantum vacuum fluctuations leads to the value which is $120$ orders of magnitude higher than the observed one.

Relations (\ref{A}), (\ref{lambda}), (\ref{M=N2m}) and (\ref{Edark}) allow us to estimate the total action of the universe,
\begin{equation} \label{A_U}
A_U := - \frac{M c^2}{H} \approx \frac{N^3A}{2\Omega_\Lambda}~,
\end{equation}
and the number of typical particles in it:
\begin{equation} \label{N}
N \approx \left(\frac{2 \Omega_\Lambda A_U}{A} \right)^{1/3} \approx \left(\frac{\Omega_\Lambda M c^2}{\pi \hbar H} \right)^{1/3} \approx 10^{40} ~.
\end{equation}
This number is known to have appeared in a different context in the Dirac's 'large numbers' hypothesis \cite{large} and is usually considered as a manifestation of a deep connection between the physics at the subatomic and cosmological scales.

Using the estimation (\ref{N}) and the formulae (\ref{A}), (\ref{Phi}), (\ref{lambda}) and (\ref{M=N2m}), from (\ref{Edark}) we can express the value of the dark energy density in our model-universe in terms of the fundamental physical parameters:
\begin{equation}
\Omega_\Lambda = N^3\frac {2\pi \hbar H^2 G}{c^5} \approx 0.7~,
\end{equation}
which is very close to the observed value \cite{Wmap}.

Now, let us consider a little bit more realistic model-universe assuming that a part of particles of the world ensemble is charged. The universe as a whole is neutral, i.e. a half of charged particles carries positive charge $+e$ and the other half have negative charge $-e$. The number of charged particles can be roughly identified with the number of baryons $N_b$ in the universe ($N_b < N$). A simple combinatorics yields for the Newtonian gravitational energy of a single baryon which interacts with all other particles in the universe the following formula:
\begin{equation} \label{Eb}
E_{b} = \left( N_b N - \frac{N_b^2}{2}\right)~\frac {Gm^2}{R} \approx  N_b N~\frac {Gm^2}{R}~.
\end{equation}
Then, according to (\ref{M=N2m}), for the total gravitational energy of the baryon component of matter we obtain:
\begin{equation} \label{Eb|G}
E_{b|G} = \frac {N^2}{2} E_b \approx \frac 12 N_b N^3~\frac {Gm^2}{R}~.
\end{equation}
It is natural to expect that the ratio (\ref{Edark}) of the gravitational and total energy is valid also for the corresponding contributions of the baryon component:
\begin{equation} \label{E/E}
\frac{E_{b|G}}{E_{b|tot}} \approx \Omega_\Lambda~,
\end{equation}
where $E_{b|tot}$ denotes the total energy of the baryon component of the universe. Then for the baryon density in the universe we obtain the following estimation:
\begin{equation} \label{Omegab}
\Omega_b \approx \frac{E_{b|tot}-E_{b|G}}{Mc^2} \approx \frac{E_{b|G}}{Mc^2}~\frac{(1-\Omega_\Lambda)}{\Omega_\Lambda}~.
\end{equation}

Further, let us estimate the total electromagnetic energy of all $N_b$ charged particles in the model-universe (i.e. that of $N_b/2$ interacting pairs). The fact that the electric charges have two polarities, while the mass is always positive, leads to basic differences from the previous consideration of the electrically neutral matter. Namely, the universe as a whole is neutral and, in contrast to the gravitational energy, the total electromagnetic, or radiation energy of a single baryon consists of $N_b/2$ additive terms, i.e.
\begin{equation}\label{Er}
E_r \approx \frac {N_b}{2} \frac{\alpha \hbar c}{R}~,
\end{equation}
where $\alpha$ is the fine structure constant. Similar to (\ref{Eb|G}), the total gravitational energy of the radiation component of matter can be estimated as:
\begin{equation}\label{Er|G}
E_{r|G} \approx N_b N^2\frac{\alpha \hbar c}{4R}~.
\end{equation}
Using this formula and the observed value of the radiation energy density \cite{Wmap}:
\begin{equation}
\Omega_r = \frac {E_{r|G}}{\Omega_\Lambda Mc^2} \approx 5 \times 10^{-5}~,
\end{equation}
we can estimate the number of baryons in the universe:
\begin{equation}
N_b \sim 10^{39}~,
\end{equation}
which turns out to be only one order of magnitude less than the estimated total number of particles (\ref{N}) in our model-universe.

Finally, equations (\ref{Eb|G}), (\ref{E/E}), (\ref{Omegab}) and (\ref{Er|G}) yield for the ratio of the radiation and baryon densities in the universe
\begin{equation} \label{omega/omega}
\frac {\Omega_r}{\Omega_b} \approx \frac{E_{r|G}}{E_{b|G}}~\frac{\Omega_\Lambda}{(1 -\Omega_\Lambda)} \approx \frac{\alpha \hbar c}{4 N G m^2}~\frac{\Omega_\Lambda}{(1-\Omega_\Lambda)}~.
\end{equation}
From (\ref{A}), (\ref{Edark}) and (\ref{lambda}) we find
\begin{equation}
\frac{\hbar c} {NGm^2} = \frac{1}{2\pi \Omega_\Lambda}~,
\end{equation}
whence it follows:
\begin{equation}\label{alpha}
\frac {\Omega_r}{\Omega_b} \approx \frac {\alpha}{8\pi (1-\Omega_\Lambda)} = 1.1 \times 10^{-3}~,
\end{equation}
which is also  very close to the observed value.}


\sectionn{Total gravitational potential $=$ twice $\Phi_{Newton}$}

{ \fontfamily{times}\selectfont
 \noindent
In this section, using different arguments, we show that the total gravitational potential of an object in the universe is scale dependent quantity and obtain twice of its Newtonian value. Only half of this total energy can be transformed to the kinetic energy, while the other half is needed to compensate the negative vacuum energy, and thus does not affect local physics. This idea is not quite new. It was already noticed that the general relativistic deflection for a test particle with an arbitrary velocity $v$ shows that gravitational mass of the particle is \cite{Car}:
\begin{equation}
  m_g = m_i \left( 1 + \frac {v^2}{c^2}\right) = m_i + 2 \frac {E_{kin}}{c^2}
\end{equation}
In the case of photons the inertial mass $m_i = 0$ and the kinetic energy, $E_{kin}$, should be replaced by the electromagnetic energy, which can be loosely interpreted as the 'kinetic energy' of photons. It is known that nonzero components of the energy-momentum tensor for light waves propagating along the $z$ axis, for example, are \cite{La-Li}:
\begin{equation}
  T^{00} = T^{33} = T^{03}~.
\end{equation}
So if we consider a free photon with the energy:
\begin{equation}
  E = \int T^{0\nu}dS_\nu
\end{equation}
and apply to it the so-called Tolman's formula for active gravitational mass \cite{La-Li}, we will obtain
\begin{equation}
  m_g = 2 \frac {E}{c^2}~,
\end{equation}
i.e. two times bigger value than the expected one. To restore the Einstein's standard relation one needs to consider interaction of the photon with other bodies. If a photon is confined in a box with mirrors, then we have a composite body at rest. In this case, as shown in \cite{Mi-Pu}, we have to take into account a negative contribution to $m_g$ from stress in the box walls and also to perform averaging over time \cite{Car,aver}.


\subsectionn{General relativity argument}

Consider a particle with the energy $E_0$ which moves freely from its position to the cosmological horizon $R$ where it has the energy $E$. The Newtonian potential energy of the particle at the horizon is:
\begin{equation}
U = -\frac{E}{c^2} \frac{GM}{R}~.
\end{equation}
Assuming the translation invariance of the energy, one should write:
\begin{equation}
E + U = E_0~.
\end{equation}
If our universe is close to the black hole state, i.e it obeys (\ref{Phi}), we find:
\begin{equation}
E = \frac {E_0} {1 - GM/c^2R} = 2 E_0.
\end{equation}
Thus, as the particle reaches the horizon, its energy and mass are doubled. Half of the particles energy at the horizon, $E=mc^2$, is needed to compensate the negative vacuum energy loss. The physical mass of the particle still is its residual mass $m_0= E_0/c^2$, and the gravitation mass $m = 2m_0$ manifests itself as a general relativistic effect in the Einstein potential $\Phi(r) = -2Gm/r$, whose value is twice of Newton's potential.

Thus the total change in potential energy of a particle of mass $m$ in the field of any mass $M$ is equal to
\begin{equation} \label{DeltaU}
\Delta U = 2G\frac {Mm}{r}~,
\end{equation}
and not to $GMm/r$ as in the Newtonian theory. At the same time, for the change in kinetic energy of $m$ we still have the classical expression,
\begin{equation}
\Delta T = G\frac {Mm}{r}~.
\end{equation}
Half of (\ref{DeltaU}) is spent to the change of the particle's internal energy:
\begin{equation}
\Delta E_0 = G\frac {Mm}{r}~.
\end{equation}

Also, the total change in the effective gravitational potential $\Delta \Phi$ is twice the value defined from the Newtonian theory:
\begin{equation}
\Delta \Phi = -2 \Delta \Phi_{Newton} = 2 \frac {GM}{r}~.
\end{equation}
For small non-relativistic systems the Newtonian potential does not lead to mistake, since only a half of the total gravitational energy transforms to the kinetic energy:
\begin{equation}
\Delta T = \frac m2 \Delta \Phi~.
\end{equation}
However, for the relativistic cases, or cosmological distances, the Newtonian theory gives wrong results.


\subsectionn{Machian considerations}

Let us recall how relativistic formulae appear in the thermodynamic model of gravity \cite{Gog}, where the universe is modeled as the world ensemble of particles. For the total energy of a particle from the ensemble, which has the kinetic energy $T$, we write:
\begin{equation} \label{Etot}
E = E_0 + T ~.
\end{equation}
The velocity dependent parameter of inertia of this particle is defined as:
\begin{equation} \label{E/Phi}
m = - \frac{E}{\Phi} ~,
\end{equation}
where, according to (\ref{Phi}), $\Phi \sim c^2$ denotes the gravitational potential of the universe acting on the particle.

The number of particles in the world ensemble is conserved. Thus the Machian energy $E_0$ is the same for all inertial observers, i.e.
\begin{equation}
E_0 = E - T = const~,
\end{equation}
or
\begin{equation} \label{dE}
dE_0 = c^2 dm - dT = 0~.
\end{equation}
Then, using the Hamilton's definition of the velocity: $v^i = dT/dp^i$, and of the momentum: $p^i = mv^i$, the latter equation can be transformed as follows:
\begin{eqnarray} \label{cal}
c^2 dm - d E_{kin} &=& c^2 dm - v^i dp_i = \left( c^2 - v^2 \right) dm - \frac m2  dv^2  =\\
&=& m \left( c^2 - v^2 \right) d \left( \ln m\sqrt{1 - \frac{v^2}{c^2}}\right) = 0~. \nonumber
\end{eqnarray}
Consequently, the quantity
\begin{equation} \label{m0}
m_0 := \frac{m}{\gamma}~,
\end{equation}
where
\begin{equation} \label{gamma}
\gamma = \frac{1}{\sqrt{1- v^2/c^2}}
\end{equation}
is the standard Lorentz factor, is constant, and hence it can be interpreted as a mass parameter of a particle (also known as the rest mass) which is valid in any inertial frame.

Returning to the main question about the total energy of a object, let us consider a generalization of the energy balance equation (\ref{Etot}):
\begin{equation}
dE = dE_0 + dT + dU~,
\end{equation}
where we had added the term contained $U = Gm\mu /r$, which is the Newtonian gravitational energy of the particle in the field of some mass $\mu$. The calculations similar to (\ref{cal}) now lead to:
\begin{equation}\label{cal-2}
m \left( c^2 - v^2 - \frac{2G\mu}{r}\right) d \left( \ln m\sqrt{1 - \frac{v^2}{c^2}- \frac{2G\mu}{c^2r}}\right) = dE_0 - \frac{G\mu}{r}dm~.
\end{equation}

In the Newtonian approximation $dm = dE_0 = 0$ and (\ref{cal-2}) yields $m = const$ and
\begin{equation} \label{T+U}
\frac m2 dv^2 - m\frac{G\mu}{r^2}dr = 0~,
\end{equation}
which leads to the standard conservation of energy in classical physics:
\begin{equation}
\frac{mv^2}{2} + mgh = const~,
\end{equation}
where $h = \Delta r$ and $g = G\mu /r^2$.

In more general case, when we introduce the standard definition of the rest mass:
\begin{equation} \label{m_0}
 m_0 = m\sqrt{1 - \frac{v^2}{c^2}- \frac{2G\mu}{c^2r}}= const~,
\end{equation}
from (\ref{cal-2}) it is clear that the rest energy of the particle is not constant, but
\begin{equation} \label{E_0}
dE_0 = \frac {G\mu}{r}dm~.
\end{equation}
This means that some part of the gravitational energy of the object $\mu$, in addition to forming the potential energy for $m$ that compensates the change in the particle's kinetic energy (\ref{T+U}), is spent to change the energy of the world ensemble, which does not affect the particle's local dynamics. Thus the effective gravitational potential of $\mu$ exceeds its Newton value. To estimate this extra part of the potential in the case of the whole world ensemble:
\begin{equation}
\mu \rightarrow M~, ~~~~~r \rightarrow R~, ~~~~c^2 \sim \frac{2GM}{R}~,
\end{equation}
let us consider the non-relativistic case: $v^2/c^2 \ll 1$. Then from (\ref{m_0}) we find:
\begin{equation}
\frac{GM}{R}dm \approx -\frac{GmM}{R^2}dR~,
\end{equation}
and for the total change of the Machian energy of a particle, $U = GmN/R$, we obtain
\begin{equation}
dU = \frac{GM}{R}dm - \frac{GmM}{R^2}dR \approx \frac{2GmM}{R^2}dR~,
\end{equation}
which is twice the Newtonian value. The Newtonian value is restored if one assumes $dm = 0$.


\subsectionn{Thermodynamic explanation}

Let us study the appearance of the factor two in the expressions of the gravitational potentials of the ensemble of massive particles (\ref{Phi}) in the thermodynamic language. In the thermodynamic approach the source of the rest energy of a particle, $E_0$, is its gravitational interaction with the world ensemble. The energy of distant particles can be described as the heat $Q$, so that
\begin{equation} \label{dE_0}
dE_0 = dQ = T dS~,
\end{equation}
where $T$ and $S$ denote the temperature and the entropy of the world ensemble.

In the thermodynamic model of gravity the inertial frames correspond to the thermodynamic equilibrium when $T = const$. The integration of (\ref{dE_0}) at constant $T$ will give
\begin{equation} \label{E_T}
E_T = TS~.
\end{equation}
This situation, when the temperature is constant in spite of the heat transfer (i.e. we neglect the energy of vacuum heating) corresponds to the Newtonian approximation, and we can write:
\begin{equation}
TS = m\frac {GM}{R} ~.
\end{equation}

On the other hand, one can take into account the energy transfer for the whole ensemble and can use the relations from the Schwarzschild black hole thermodynamics:
\begin{equation}
T = T (E) = \frac {1}{2 C E}~, ~~~~~ S = C E^2~,
\end{equation}
where the constant $C = 4\pi k_BG/\hbar c^5$ is expressed as a combination of fundamental physical constants. Then the integration of (\ref{dE_0}) will lead to
\begin{equation} \label{E=}
E  = 2TS = m \frac {2MG}{R}~.
\end{equation}

Thus the expression (\ref{dE_0}) arises when $T$ is held constant, while (\ref{E=}) arises when $T$ is treated as a specified function of $E$. This brings to mind the analogy with the canonical ensemble (with the constant temperature $T$) and the micro-canonical ensemble (in which the energy $E$ is held constant).

In the case of Einstein's gravity the thermodynamic expression similar to (\ref{E=}), in the context of a general horizon, was considered in \cite{Pad}. The energy (\ref{dE_0}), on the other hand, arises whenever one considers transfer of energy across any null surface, as viewed by a local Rindler observer. This expression is applicable to the cases in which the injected energy is not considered as a part of a self-gravitating system and one can keep the temperature of the horizon constant in spite of the injection of the energy.


\subsectionn{Renormalization group-like analysis}

In particle physics the vacuum energy itself is unobservable, only the quantum fluctuations have a physical meaning. As first suggested in \cite{Zel}, the vacuum energy can be given by the gravitational energy of the virtual particle-antiparticle pairs which are continuously created and annihilated in the vacuum. This Newtonian energy reads $Gm^2(r)/r$, where $m(r)$ is the effective mass at the scale $r$. In the thermodynamic model the expression for the mass function $m(r)$ is related to the structure of the universe, and the mass of a particle, $m_1$, is an explicitly scale dependent quantity, as it is already the case for several quantities in quantum field theory. For example, the electric charge is known to increase when the length-scale decreases below the Compton length of the electron, as a result of vacuum polarization by virtual particle pairs, and the 'bare' value of charge is much higher than its Bohr value \cite{It-Zu}. For the mass function in the thermodynamic model we have the opposite picture, because of the influence of distant particles the value of mass should be higher for larger scales. However, the mechanism is similar and we can write the renormalization group equation for the energy density parameters:
\begin{equation} \label{renorm}
\frac{d \Omega_i}{d \ln r} = f(\Omega_i) ~,
\end{equation}
where $i$ labels massive objects in the universe. The function $f(\Omega_i)$ is unknown, but since $\Omega_i=m_i/M\ll 1$, it may be expanded to the first order about the origin:
\begin{equation}
f(\Omega_i) \approx -\alpha (\beta_i -  \Omega_i) ~,
\end{equation}
where $\alpha$ and $\beta_i$ are some constants. Then equation (\ref{renorm}) can be solved:
\begin{equation} \label{Omega-1}
\Omega_i = \beta_i\left[ 1+ \left(\frac{r}{R}\right)^\alpha \right]~,
\end{equation}
where the integration constant $R$ is taken to be of the order of the horizon radius.

Equation (\ref{Omega-1}) tells us that for small scales $r\ll R$ we measure the mass of a particle:
\begin{equation} \label{m-1}
m_i(r\ll R) = M\beta_i~,
\end{equation}
and at the horizon scales the mass is twice of this value:
\begin{equation}
m_i(r=R) = 2 M\beta_i~.
\end{equation} }


\sectionn{Newton's second law and the Machian mass}

{ \fontfamily{times}\selectfont
 \noindent
 We want to show that within the thermodynamic model \cite{Gog} it is possible to describe the acceleration as a relative quantity in the spirit of (\ref{F}). Since we describe the universe as the ensemble of particles let us use the Einstein-Infeld-Hoffmann equations for the gravitationally interacting $N$ classical objects \cite{EIH}. These equations are based on the Lagrangian \cite{Fock}:
\begin{eqnarray}\label{L}
L &=& \sum_a \frac{m_av^2_a}{2} + \frac{1}{2} \sum_a \sum_{b\neq a}  G\frac{m_a m_b}{r_{ab}} + O\left(\frac{v^2}{c^2}, G^2, ...\right) +\nonumber \\
&+& \frac{1}{4c^2} \sum_a \sum_{b\neq a}  G\frac {m_a m_b}{r_{ab}} \left[ 3\left(v_a^2 + v_b^2\right) - 7(\vec{v}_a\cdot\vec{v}_b) - \frac{(\vec{v}_a \cdot \vec{r}_{ab}) (\vec{v}_b \cdot \vec{r}_{ab})}{r_{ab}^2} \right]~,
\end{eqnarray}
were $\vec{r}_{ab} =\vec{r}_b - \vec{r}_a$ is the radius-vector from particle $a$ to particle $b$ ($r_{ab} = |\vec{r}_{ab}|$). Here the gravitational interaction between pairs is modeled by the classical Newton potential:
\begin{equation}
U_{ab}= G \frac {m_a m_b}{r_{ab}}~.
\end{equation}
The equation of motion for a particle from the world ensemble, which we label by $1$, is given by the Euler-Lagrange equation:
\begin{equation}
\frac{d\vec{p}_1}{dt}= \vec{F}_1 ~,
\end{equation}
where the generalized momentum can be found from (\ref{L}) to have the form:
\begin{equation}\label{p1}
\vec{p}_1 = \frac{\partial L}{\partial \vec{v}_1} = m_1 \vec{v}_1 + \frac{G m_1}{2 c^2} \sum_{b=2}^N \frac{m_b}{r_{1b}} \left[ 6\vec{v}_1 - 7\vec{v}_b - \frac{\vec{r}_{1b} (\vec{v}_b \cdot \vec{r}_{1b}) }{r_{1b}^2}\right]~.
\end{equation}
For simplicity we have neglected the term $O$ in (\ref{L}), which contains the relativistic correction to the classical kinetic energy $\sim (v/c)^2$ and the higher order corrections to the gravitational interaction $\sim G^2$.

We see from (\ref{L}) that, while the forces arising from $\vec{F}_1 = \partial L/\partial \vec{r}_1$ decrease at least as $1/r^2$, the inertial terms in (\ref{p1}) decrease only as $1/r$. Consequently the inertia has an intrinsically non-local nature and is intimately connected to cosmology.

Assume now that a selected particle is at the origin in a homogeneous isotropic expanding universe of density $\rho$ and with the Hubble parameter $H$. The particle has the position $\vec{r}$ and the velocity
\begin{equation}
\vec{v} = H \vec{r} + \vec{u}~,
\end{equation}
were $\vec{u}$ is an overall velocity of the universe relative to the origin. Other particles, $m_b$, we replace by the mass element,
\begin{equation} \label{2rho}
\sum_{b=2}^N m_b \rightarrow 2\rho d V~,
\end{equation}
where $V = 4\pi R^3/3$ is the volume of the universe. Note the appearance of the extra factor $2$ in this formula, in agreement with the results of the Section $3$.

We then replace the sum in (\ref{p1}) by the integral:
\begin{equation}\label{p1-2}
\vec{p}_1 = m_1 \vec{v}_1 + \frac{G m_1}{c^2} \int \frac{\rho}{r} \left\{ 6\vec{v}_1 - 7(H \vec{r} + \vec{u}) - \frac{\vec{r} \left[ H r^2 + (\vec{u} \cdot \vec{r})\right] }{r^2}\right\} dV~.
\end{equation}
We need to calculate the integrals in this expression for the spherical volume, $d V = r^2 \sin\theta d r d\varphi d\theta$, over the visible universe, $R=c/H$.

Without the loss of generality we assume:
\begin{equation}
\vec{u} = u \vec{e}_z~.
\end{equation}
Since $\vec{r} = r \left( \sin \theta \vec{e}_\varphi + \cos\theta \vec{e}_z \right)$, where $\vec{e}_\varphi = \cos\varphi \vec{e}_x + \sin\varphi \vec{e}_y$, the scalar product term in (\ref{p1-2}) becomes:
\begin{equation}
\vec{r} (\vec{u} \cdot \vec{r}) = r^2 u \left(\sin\theta \vec{e}_\varphi + \cos\theta \vec{e}_z \right) \cos\theta~.
\end{equation}
The terms involving $H$ in (\ref{p1-2}) are multiplied by $\vec{r}$ and the integrations over the sphere of radius $R$ make them vanish for symmetry reasons. The term multiplying $\vec{e}_\varphi$ also vanishes since nothing depends on the angle $\varphi$. Then we need to calculate only two different integrals:
\begin{eqnarray}
\int \frac{\rho~ d V}{r} &=& 4\pi \rho \int_0^R  r d r = 2\pi \rho R^2~, \nonumber \\
\int \frac{\rho \cos^2\theta ~d V}{r} &=& 2\pi \rho \int_0^R r d r \int_0^{\pi} \cos^2\theta ~\sin\theta ~d\theta = \frac{2}{3}\pi \rho R^2~.
\end{eqnarray}
Inserting these results into the expression of the momentum (\ref{p1-2}), we find:
\begin{equation}\label{p1-3}
\vec{p}_1 = m_1 \left[ \left( 1 + \frac{9}{2} \Omega \right)\vec{v}_1 - \left( \frac{11}{2} \Omega \right) \vec{u} \right]= \frac{11}{2}m_1\left(\vec{v}_1 - \vec{u} \right)~.
\end{equation}
Here we have inserted the cosmological density parameter of the flat cosmological model
\begin{equation} \label{Omega-crit}
\Omega = \frac {\rho}{\rho_c} = 1~.
\end{equation}

The formula (\ref{p1-3}) differs by the factor two from the result of \cite{Ess} and \cite{Mol}, where the unrealistic condition $\Omega =2$ was introduced to obtain the correct answer. The reason of discrepancy is our extrapolation (\ref{2rho}), which is based on the arguments considered in Section $3$. Our result is also in a qualitative agreement with other investigations which find that Mach's principle requires the density parameter of flat cosmological model \cite{Omega=1}.

Now note that for the considered Einstein-Infeld-Hoffmann ensemble of $N$ classical particles the parameter $\rho$ should correspond only to baryonic matter whose density is much less than $\rho_c$. However, since we assume that the ratios like (\ref{Edark}) are valid also for the corresponding contributions of the baryon component (\ref{E/E}), we expect that the structure of (\ref{p1-3}) will survive if we use the relation (\ref{Omega-crit}). Besides, in this case the results of calculations will not depend on the cosmological epoch, since $\Omega$ of the flat cosmological model is time independent. At the same time, in order to compensate the mass exceed because of the assumption (\ref{Omega-crit}), we need to "renormalize" the "bare" mass $m_1$ of the simple model (\ref{L}):
\begin{equation} \label{m1}
m_1 \approx \Omega_M m~,
\end{equation}
to the actual mass of the particle $m$.

Returning to (\ref{p1}) we note that, as it is seen from the relations (\ref{p1-3}) and (\ref{m1}), if
\begin{equation}
\Omega_M = \frac {2}{11} \approx 0.2 ~,
\end{equation}
then one can explain the inertia of a particle by its gravitational interactions with the whole universe:
\begin{equation}
 \vec{p}_1 \approx m (\vec{v}_1 - \vec{u} ) ~~~~~ \Rightarrow ~~~~~ \vec{F}=\frac{d\vec{p}_1}{dt} \approx m (\vec{a}_1 - \vec{a}_u )~,
\end{equation}
and conclude that the acceleration in Newton's second law can be considered as a relative quantity with respect to the universe.
}


\sectionn{Conclusions}

{ \fontfamily{times}\selectfont
 \noindent
In this paper we assumed the existence of the relations between the fundamental physical constants and the cosmological parameters using the thermodynamic approach of \cite{Gog}. The dark energy is identified with the energy of collective gravitational interactions of all particles in the universe, which is not taken into account in the standard treatments. The obtained values for the radiation, baryon and dark energy densities are close to the current cosmological observations. It is shown that the total energy, or mass, of any object in the universe is a scale dependant quantity and obtains twice of its Newtonian value for the whole world ensemble. Using these results it was found that a precise formulation of the Mach's principle can be consistent with the Einstein-Infeld-Hoffmann approximation to general relativity in the case of flat cosmological model.
}


 {\color{myaqua}

 \vskip 6mm

 \noindent\Large\bf Acknowledgments}

 \vskip 3mm

{ \fontfamily{times}\selectfont
 \noindent
 This research is supported by the grant of Shota Rustaveli National Science Foundation $\#{\rm DI}/8/6-100/12$ and its publication is supported by Volkswagenstiftung under the contract $\#{\rm 86260}$.
 

 {\color{myaqua}

}}

\end{document}